\documentclass[runningheads]{llncs}

\usepackage[american]{babel}
\usepackage[utf8]{inputenc}
\usepackage{listings}
\usepackage{times}
\usepackage{cite}
\usepackage{units}
\usepackage{setspace}
\usepackage{graphicx}
\usepackage{url}
\usepackage{epsfig}
\usepackage{booktabs}
\usepackage{bm}
\usepackage{comment}
\usepackage{color}
\usepackage{xspace}
\usepackage{epstopdf}
\usepackage{xcolor}
\usepackage{float}
\usepackage{framed}
\usepackage{array}
\usepackage{makecell}
\usepackage{amssymb}

\begin{document}

\title{Challenges and Directions for Authentication in Pervasive Computing}

\author{Artur Souza  \and Antônio A.F. Loureiro \and Leonardo B. Oliveira\thanks{Currently, a Visiting Associate Professor at Stanford University.}}
\authorrunning{Artur Souza et al.}

\institute{Universidade Federal de Minas Gerais, Brazil \email{\{arturluis,loureiro,leob\}@dcc.ufmg.br}}

\maketitle

\begin{abstract}
    We quickly approach a “pervasive future” where pervasive computing is the norm. In this scenario, humans are surrounded by a multitude of heterogeneous devices that assist them in almost every aspect of their daily routines. The realization of this future demands strong authentication guarantees to ensure that these devices are not abused and that their users are not endangered. However, providing authentication for these systems is a challenging task due to the high heterogeneity of pervasive computing applications. This heterogeneity makes it unfeasible to propose a single authentication solution for all of the pervasive computing applications. In this paper, we review several pervasive application scenarios and promising authentication methods for each. To do this, we first identify the key characteristics of each pervasive application scenario. Then, we review the strengths and weaknesses of prominent authentication methods from the literature. Finally, we identify which authentication methods are well suited for each application scenario based on the identified characteristics. Our goal is to provide promising directions to be explored for authentication in each of these scenarios. 
\end{abstract}

\section{Introduction} \label{sec:intro}

\begin{figure}[htb]
    \begin{center}
    \includegraphics[width=0.9\textwidth]{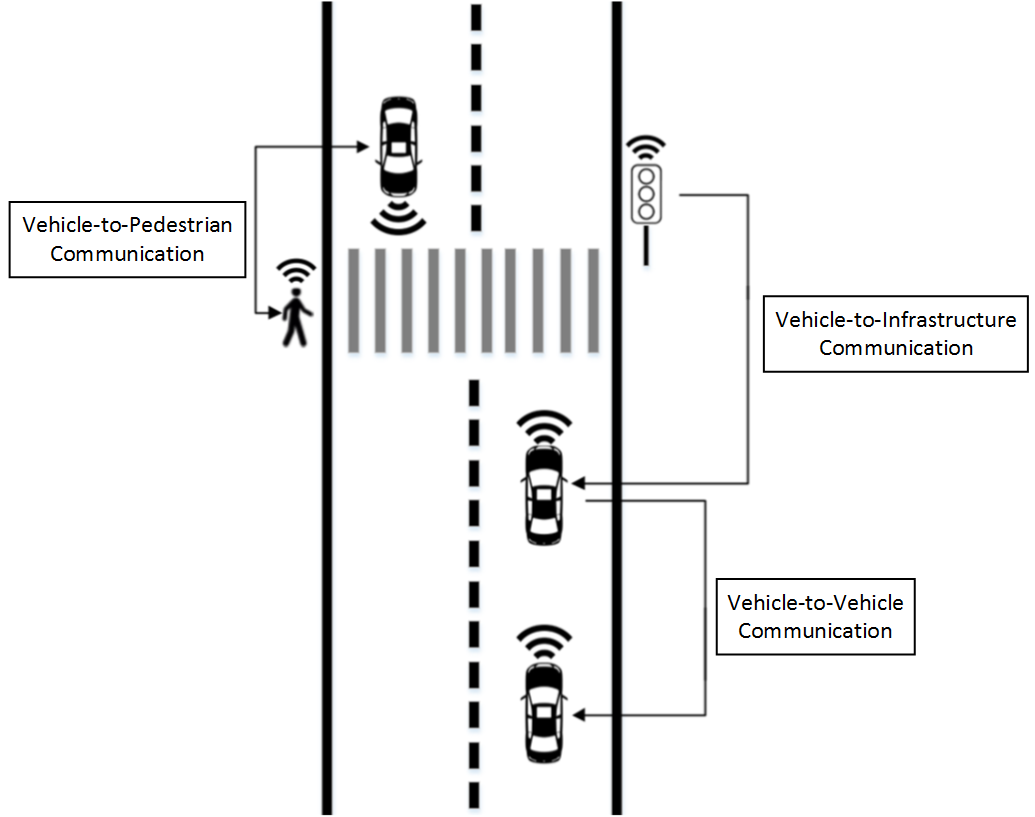}
    \caption{Example of pervasive computing in the context of vehicular networks.}
    \label{fig:pervasive_example}
    \end{center}
\end{figure}

Pervasive computing, or ubiquitous computing, denotes a paradigm where users are surrounded by computing elements and sensors attached to all sorts of everyday objects~\cite{satyanarayanan2001pervasive}. This enables a series of new and ubiquitous applications that impact every aspect of users’ day-to-day routine~\cite{Atzori10}. 
Figure~\ref{fig:pervasive_example} shows an example of pervasive computing in the context of vehicular networks. In this case, the computing elements (vehicles, road infrastructure, and even the user's personal devices) interoperate to improve traffic efficiency and safety, sometimes even without the users noticing (c.f. Section~\ref{sec:scen.v2x}).

This ``pervasive future'', however, brings forth a series of security concerns~\cite{Jing14, Kaur17game}. In this scenario, computing elements are ubiquitous and often track every aspect of users daily routines. This means devices have access to vast amounts of sensitive information about its users, which, in turn, raises privacy concerns~\cite{Souza17Nomadikey, galdi2017secure}. Besides, the rapidly growing number of devices in these systems mean these networks will be massively larger than other networks, meaning adversaries have several more targets to exploit. A good example on the risk of leaving these systems unprotected is the recent Mirai botnet attack~\cite{antonakakis2017understanding}.


Notably, authentication is a critical security property for pervasive computing systems. Authentication ensures that devices can verify the header (source authentication) and content (data authentication) of messages received~\cite{Stinson02}. This forms the basis for secure communication channels, since it allows devices to trust messages exchanged. In a pervasive system, authentication allows devices to safely communicate and cooperate to provide their intended applications. 

Providing authentication in this complex pervasive scenario, however, is a challenging task~\cite{ouechtati2017trust}. Most of all, because of the heterogeneity of the network. Providing a single authentication solution that fits the needs of each of these applications is an unfeasible task. Most notably, the heterogeneity and peculiarities of these systems render traditional solutions inadequate and, therefore, call for novel solutions tailored for these systems.

In this paper, we perform a study of different pervasive application scenarios and authentication solutions. We first review several pervasive computing applications, highlighting their main particularities and their impact on security solutions. We then review several authentication solutions that are promising for the devices and applications often found in pervasive computing. Finally, we identify promising combinations of application scenarios and authentication solutions, as well as open challenges and limitations of current approaches for each specific scenario.

This paper is organized as follows. Section~\ref{sec:scen} presents the different applications we consider in this paper. Section~\ref{sec:method} presents existing authentication strategies. Section~\ref{sec:applied} identifies promising authentication directions for each application scenario. Finally, in Section~\ref{sec:conc} we present our concluding remarks.

\section{Pervasive Computing Scenarios} \label{sec:scen}

The essence of pervasive computing is that computing elements are coupled to everyday objects in order to allow them to communicate and interoperate to provide applications~\cite{satyanarayanan2001pervasive}. Pervasive computing enables applications for almost every aspect of users' daily routines. In this section, we present the most prominent pervasive application scenarios and review their particularities for security solutions. 

\subsection{Wireless Sensor Networks}

Wireless Sensors Networks (WSNs) comprises a broad spectrum of pervasive computing where the
computing elements are sensors distributed in an wide area to monitor physical and environmental conditions~\cite{Akyldiz10}. These sensors are often linked to a base station to which they report the sensed data~\cite{Karlof04, Akkaya05}. Applications of WSNs vary greatly based on what is being sensed and where, for instance, sensors may be deployed in crops to monitor humidity and soil conditions or throughout a city to monitor its levels of air pollution~\cite{Arampatzis05}.

There are several limitations to the development of authentication solutions for WSNs~\cite{perrig06seven}. First, sensors are often severely resource-constrained~\cite{perrig06seven} and have no sustainable power source~\cite{Ye02}. This means authentication solutions for WSNs must be lightweight, so that computations can be done in a timely manner and  with minimal impact to the sensors lifespan. Besides, sensors are often deployed in unsupervised environments, which makes it easier for adversaries to physically access the device and attempt to physically extract information (including cryptographic material) from the device~\cite{Perrig02spins}. On the other hand, one powerful advantage of WSNs is that sensors are all bootstrapped in the base station before being deployed~\cite{Akkaya05}. This facilitates the deployment of cryptographic material inside the sensor.

\subsection{Vehicle-to-Everything} \label{sec:scen.v2x}

As its name suggests, Vehicle-to-Everything (V2X) comprises the communication between vehicles and any other interlocutor~\cite{amadeo2016}. V2X encompasses communication patterns like Vehicle-to-Vehicle (V2V or VANETs), Vehicle-to-Infrastructure or Vehicle-to-Roadside (V2I or V2R), Vehicle-to-Pedestrian (V2P), and Vehicle-to-Device (V2D). V2X applications can be classified into (i) safety applications, (ii) entertainment applications, (iii) efficiency applications~\cite{jakubiak08}. Examples of the former include forward collision warning, electronic emergency brake lights, and wrong way driving warning~\cite{jakubiak08}. Entertainment applications include media streaming, car to car messaging, and local touristic information~\cite{qian2008design}. The latter refers to applications like congestion control, improved location systems, and improved route planning~\cite{Bauza2010road}. Finally, V2X is also the cornerstone for Intelligent Transportation Systems (ITS), which are expected to greatly improve public transportation services~\cite{karagiannis11}.

There are several complications to the design of authentication solutions for V2X~\cite{weiss2011, Raya07}. The first is the vehicles' high-speed mobility, which means applications are very time sensitive and, therefore, authentication must incur in minimal time overhead~\cite{Raya07}. The second is the network's massive size, VANETs are likely to be one of the largest mobile ad hoc networks~\cite{Raya07}, this imposes burdens in the deployment and management of cryptosystems and incurs in computational overheads as it means vehicles will have to authenticate several other nodes in the network~\cite{kargl2015wireless}. Third, V2X suffer from intermittent connectivity as vehicles' will sometimes move to areas without connectivity like highways or rural areas~\cite{Raya07}, which means solutions cannot rely on a central entity available at all times to manage the cryptosystem. Finally, privacy is a major concern in V2X as information about the driver's identity or location may facilitate the action of adversaries and put the driver at risk~\cite{Forster16}. Despite the above complications, V2X networks present some advantages over other scenarios. The first being that vehicles carry a rechargeable battery, meaning energy overheads are not a serious concern. Another advantage is that vehicles are not severely resource-constrained, meaning they can afford to run more expensive cryptographic algorithms, especially for applications that are not time sensitive~\cite{Raya07}. 

\subsection{Smart Homes} \label{sec:scen.smarthomes}

As its name suggests, Smart Homes refers to scenarios where devices are coupled to common household objects, like existing smart home appliances. The main goal of smart home applications is to provide functionalities to automate or facilitate aspects of the user's domestic routine. Example of applications include remotely controlled smart lightbulbs\footnote{http://www2.meethue.com/} and smart thermostats~\footnote{https://nest.com/thermostat/meet-nest-thermostat/}. Users have a very active role in smart home environments. Most of the times, users are directly responsible for managing and activating the applications (e.g., turning on the lightbulbs or adjusting the room's temperature). This contrasts with other scenarios, like V2X, where devices often act on their own (e.g. a vehicle decides to brake when it anticipates an accident). 

The direct involvement of the user with the IoT devices, especially requiring that users set up the devices, directly impacts authentication solutions. That is because authentication solutions for smart homes must then be seamless and easy to set up. Otherwise, they risk annoying the users to the point where users simply drop or undermine security for the sake of easier use~\cite{Souza17Nomadikey}. At the same time, however, the involvement of users also enables new functionalities or even communication channels as humans may participate in the authentication process~\cite{Neto16}. 

Another key component in smart home environments is the heterogeneity of devices. Domestic IoT devices may range from powerful smartphones and tablets to resource-constrained smart lightbulbs and door locks. This heterogeneity complicates the proposal of novel authentication solutions, as the solutions must be simultaneously fit for widely different devices. On the other hand, the presence of powerful devices in the network facilitate authentication solutions, since expensive operations can sometimes be offloaded to these devices~\cite{Neto16}.

Finally, there are a couple of other advantages for authentication solutions in smart homes. First, devices are directly connected to a power source, which makes energy efficiency less of a concern, although still desirable. Second, smart homes are often controlled environments with difficult access for adversaries, which means physical attacks are less likely.

\subsection{Wearables and Implanted Medical Devices}

Wearable and Implanted Medical Devices (IMDs) are devices that are attached to the user's body~\cite{Rostami13, Wei14}. Although there is a fundamental difference in purpose between Wearables and IMDs, we grouped these two scenarios together due to their similarities for authentication. The core idea in this scenario is that devices are attached to users bodies and they attempt to insert the user himself as a component in the pervasive paradigm~\cite{Wei14}. These devices also act as body sensors, monitoring physical conditions like body temperature, movement, and health conditions (e.g. heartbeats)~\cite{Rostami13}.

Because of the amount of sensitive information stored in these devices, privacy is a paramount concern~\cite{Raij11}. Illegitimate access to these devices allow adversaries to access the user's private information (e.g., emails, messages), body conditions (e.g., temperature, movement), and even monitor medical conditions (e.g., presence of pacemakers)~\cite{Raij11, Rostami13}. At the same time, these devices often have low computational power and rely on other, more powerful, devices to perform computations~\cite{DiPietro03}. 

There are a few similarities between Wearables/IMDs and previous application scenarios. For instance, similar to sensors in WSNs, these devices have low computational power and they also have no sustainable energy source, although wearable devices can be more easily recharged. Besides, as it happens in smart home environments, these devices sometimes require direct user involvement, which concurrently brings new challenges and possibilities for these applications. These devices can also rely on the presence of other powerful devices (to which they report data) to perform heavier operations and are often in very controlled environments (worn by the users or attached to their bodies), which complicates the action of adversaries.

Unlike previous applications, however, Wearables and IMDs are constantly in contact with the user's body. This creates possible new venus for authentication, based on the user's physical information. For instance, a pacemaker can use information about the heartbeat to authenticate~\cite{Rostami13}, while fitness trackers can use gait recognition~\cite{Derawi10} for authentication. 

\subsection{Cyber-Physical-Human Systems}

Cyber-Physical-Human Systems (CPHSs) is a rendezvous between cyber systems, the physical world, and human beings. Cyber-Physical-Systems (CPSs) refer to the integration of the cyber and physical worlds~\cite{lee2008cyber}. CPHSs, in turn, put people in the loop by making them not only a user of CPSs but also part of them~\cite{Wu17PCS}. CPHSs applications are highly diverse, including cyber-physical medical systems, critical infrastructure control, workspace automation, among many others~\cite{cardenas2008secure}. The broad definition of CPHS may even include vehicular systems or wearable devices~\cite{cardenas2008secure, lee2008cyber}, although in this work we consider these as different applications.

The high diversity of CPS, and consequently CPHS, make the conception of authentication solutions a challenge since resources in one context may not be available in others. However, there are two characteristics that are staple for CPHS. First of all, humans are directly involved in CPHS applications, for instance, the workers in a workspace or the doctors in cyber-physical systems~\cite{schirner2013future}. Further, CPHS aim at integrating the cyber and physical worlds, meaning environmental conditions will always be a central piece of these systems. The integration of characteristics of the physical world in the authentication is a promising venue for CPHS.

\begin{table}[t]
  \caption{Promising authentication methods for IoT application scenarios.}
  \label{tab:scenarios}
  \centering
  \resizebox{0.9\textwidth}{!}{%
  \begin{tabular}{c|c|c|c|c|c}
    \multicolumn{1}{c|}{} & {\bf \makecell[tc]{Device \\ Heterogeneity}} & {\bf \makecell[tc]{Power \\ Supply}} & {\bf \makecell[tc]{Controlled \\ Environment}}  & {\bf \makecell[tc]{User \\ Interaction}}  & {\bf \makecell[tc]{Network \\ Size}}  \\ \hline
    WSN                & \checkmark    &              &             &            & \checkmark \\ \hline
    V2X                &               & \checkmark   &             &            & \checkmark \\ \hline
    Smart Homes        & \checkmark    & \checkmark   & \checkmark  & \checkmark &            \\ \hline
    Wearables / IMD    & \checkmark    & \checkmark   & \checkmark  & \checkmark &            \\ \hline
    CPHS               & \checkmark    &              &             & \checkmark &            \\ 
  \end{tabular}%
  }
\end{table}

\subsection{Summary} \label{sec:scen.summary}

Table~\ref{tab:scenarios} shows a summary of the common constraints for authentication solutions and pervasive computing applications. We add checkmarks to the constraints that affect each application scenario. We note that, since there is great variety in CPHS applications, there might be exceptions to each constraint, but we focus on the constraints that are more frequent or more impactful. For instance, some CPHS may not have to worry about resource-constrained devices, but ideally the authentication solution would support these devices for when they are present. Besides, there are some constraints that are unique to certain application scenarios and, thus, we do not add them to the table. This does not mean these constraints are any less important, for instance, mobility is a key factor in V2X (and maybe wearables/IMD), but is not common in any of the other application scenarios.

The columns in the table refer to the most frequent constraints from our analysis in the preceding subsections. {\em Device Heterogeneity} refers to the presence of devices with greatly varied computing resources. Usually, this means interoperation between resourceful and resource-constrained devices. One classic example is the interaction between constrained sensors and resourceful base stations in WSNs. {\em Power Supply} means devices are usually connected to a power source and, thus, energy efficiency is not necessarily a concern (although often desirable). For instance, appliances in smart homes will likely be constantly connected to power outlets. {\em Controlled Environment} means devices are deployed in controlled environments, where it is harder for adversaries to physically access devices without getting caught. This is the case for Wearables and especially IMDs, which are almost always connected to the user's body. {\em User Interaction} means human users will directly interact with the devices. For instance, users may be required to manually bootstrap devices in smart homes, thus, security solutions should have user experience in mind (c.f. Section~\ref{sec:scen.smarthomes}). Finally, {\em Network Size} means the networks may contain a massive number of devices interoperating and, thus, authentication solutions have to be scalable. For instance, vehicles in V2X may easily surpass the millions in a single country (e.g., until 2017, there were almost 300 millions vehicles in the United States\footnote{https://www.statista.com/statistics/183505/number-of-vehicles-in-the-united-states-since-1990/}), all managed in the top-level by some sort of government agency.


\section{Authentication Methods} \label{sec:method}

There are several different methods for achieving authentication of users and
data. Each of these methods has its own advantages and disadvantages, which
makes it more or less suitable for each application scenario. In this section,
we review some of the most prominent authentication methods in the literature.

\subsection{Digital Signature Algorithm}

As a baseline for comparison for the other authentication mechanisms, we first present the traditional digital signature algorithm (DSA)~\cite{menezes1996handbook}. The DSA is one of the most widespread authentication solutions in traditional authentication solutions. For instance, DSA is supported by the TLS/SSL security protocols. 

Despite its widespread use, the combination of digital certificates and the DSA incurs high computational overheads~\cite{Stinson02, oliveira2011secure}. The DSA is expensive in time and storage (large signature and key sizes), while digital certificates incur in high communication, storage, and management overheads, especially in big networks~\cite{oliveira2011secure}. The cost of this solution renders it inadequate for most pervasive computing applications~\cite{Stinson02}. Still, a few authentication solutions using DSA have been proposed~\cite{Raya07}.

\subsection{Elliptic Curve Digital Signature Algorithm}

The Elliptic Curve Digital Signature Algorithm (ECDSA) is a more viable alternative to the traditional DSA~\cite{johnson2001elliptic}. The ECDSA is much more efficient in both storage and computation than the traditional DSA as it requires smaller parameters to sign and the signature generation is significantly faster~\cite{oliveira2011secure}. The ECDSA still requires a certification method, like traditional digital certificates. 

To avoid the high costs of digital certificates, Brown {\em et al.} have proposed a novel certification scheme for ECDSA called {\em Implicit certificates}~\cite{Brown01}. The main idea behind implicit certificates is that the user's public key and its certificate are combined in a single element, called the public key reconstruction data. Implicit certificates are considerably smaller and faster to verify than traditional certificates~\cite{Brown01}. 

\subsection{Short Signatures}

The Boneh-Lynn-Shacham short signature scheme (BLS)~\cite{boneh2001short} is a certificate-based signature scheme that generates signatures that are much shorter than DSA and ECDSA signatures~\cite{oliveira2011secure}. The small size of the signatures generated is the main strength of BLS. Another interesting aspect of BLS is that its computational overhead is asymmetric: verifying signatures is much more expensive than generating them. This makes BLS well-suited for applications where signatures are generated by resource-constrained devices but verified by powerful devices (e.g. Section~\ref{sec:applied.wsn}). These signatures are still certificate-based, however, meaning they suffer from the complexity and high overhead of digital certificates.

\subsection{Identity-Based Cryptography}

The idea behind Identity-Based Cryptography (IBC) was first proposed by Shamir~\cite{shamir84}, but was only made possible with the advent of pairing-based cryptography (PBC)~\cite{boneh01}. The key idea of IBC is that user's keys can be directly derived directly from the user's identity in the system (e.g., his email address). This means the user's key is intrinsically bound to the user, which, in turn, dismisses the need for other certification methods like digital or implicit certificates. Since IBC dismisses certification, the cryptosystem incurs lower storage and communication overheads and simplifies key management.

IBC, however, is not a panacea. In IBC, private keys are not generated by their respective owners but rather by a Private Key Generator (PKGs), a central entity in the cryptosystem. This means the PKG must be trustworthy, since it is able to impersonate any user in the system~\cite{chatterjee2011identity}. This is the well-known key escrow problem of identity-based systems. Besides, the centralized key generation creates a big vulnerability in the system if the PKG is compromised. At last, bilinear pairings, the mathematical foundation of IBC, are computationally expensive to compute, consequenly, so are cryptographic operations in IBC.


\subsection{Certificateless Schemes}

The Certificateless scheme~\cite{Riyami03} was proposed as a middle ground between IBC and traditional Public-Key Infrastructure (PKI). The cryptosystem is similar to IBC in that it dismisses certification schemes. But, it avoids the key escrow problem by splitting the key generation between the user and a central authority in the cryptosystem. The main advantage of Certificateless is that keys can be easily shared and verified, without the key escrow problem. One drawback of the Certificateless approach, however, is that public keys can no longer be easily derived from the user's identity like in IBC. Instead, public keys must be explicitly made public by their owners. Besides, Certificateless is also based on expensive bilinear pairings~\cite{Neto16}.

\subsection{Attribute-Based Cryptography}

Attribute-Based Cryptography (ABC)~\cite{goyal2006attribute}, is an extension of IBC that focuses on attributes instead of identities. In ABC, users' keys are derived from a set of attributes they possess, rather than their identifier. In this setup, users no longer sign messages to prove their identities, instead, users prove they possess a set of attributes from the system. ABC is particularly well-suited for access-control policies, especially  attribute-based access control~\cite{yuan2005, ouechtati2017trust}. As an extension of IBC, however, ABC also suffers from the key escrow problem and bilinear pairings.

\subsection{Hash-Based Signatures}

Hash-Based Signatures (HBSs) are signature schemes based solely on cryptographic hash functions~\cite{butin2015real, rohde08}. HBSs have gained a lot of attention in the last decade because of their high efficiency and resistance to quantum cryptanalysis~\cite{bernstein2009introduction}. HBSs are often classified into two categories: One-Time Signature schemes (OTS)~\cite{hulsing2013w} and Multi-Time Signature schemes (MTS)~\cite{buchmann09}. As their name suggests, in OTS private keys can only be used to sign messages once and must then be renewed, while in MTS, private keys can be used to sign a large, but still limited, number of messages before being renewed. The need for constantly renewing private keys can become a burden in HBS solutions. Another disadvantage of HBSs is that they also require some form of certification, much like DSA and ECDSA. Unlike ECDSA, however, there is no cheaper certification alternatives for HBS and, thus, they rely on expensive digital certificates. 

\subsection{Group Signatures}

In group signature schemes, users sign messages to prove they are a valid member of a group, instead of proving their identity in the system~\cite{chaum1991group}. Group signatures are always anonymous in the sense that it is not possible for other users to know who generated a signature. In these schemes, a group manager is responsible for generating and distributing keys to other users in the system. The group manager is also responsible for adding or removing users from the group. Unlike the other members of the group, the group manager is able to ``open'' a signature to reveal its signer. This is used to identify and remove malicious users in the group.

Group signatures have the clear advantage of being anonymous and simple, as each user need only know the group generation and verification keys. However, there is significant overhead for the group manager to manage the devices, especially if the group is highly dynamic. Also, since all devices use the same group key, a single device that is compromised means the entire group must be redone.

\subsection{Ring Signatures}

Ring signatures are an extension of group signature schemes~\cite{rivest2001leak}. Much like in group signature schemes, users sign messages as one of the members of a group of users. However, in ring signature schemes, the group need not be truly formed and the other users need not even be aware of the group. The user may simply choose any set of users and sign a message as one of the users in the set.

There are two main differences between group signatures and ring signatures. The first is that ring signature schemes do not need group managers since the groups are formed by the user himself when generating the signature. The second is that, since there is no group manager, no one is able to trace back the generator of a signature, therefore the signatures are completely anonymous.

Ring signatures are interesting replacements for group signatures for scenarios where there is no suitable candidate for the group manager or when groups are difficult to form. One drawback of ring signatures, however, is that they require a previous set up of public/private key pairs and the efficiency of the ring signature depends on the underlying cryptosystem~\cite{rivest2001leak}.

\subsection{Mesh Signatures}

Mesh signatures are an extension of Ring Signatures~\cite{boyen2007mesh}. Similar to its precursor, mesh signatures allow users to sign a message anonymously, providing only a list of possible signers. Mesh signatures, however, possess two differences when compared to ring signatures. First, mesh signatures allow any individual to be included in the list of possible signers, even if the user real public key of the user is unknown. Second, mesh signatures provide threshold signing. That is, messages may be signed as a subset of members of the group of possible signers. 

One drawback of mesh signatures, however, is that it enables users to collude to generate signatures that each of the users alone would not be able to generate~\cite{Maji11}. This could allow users to collude to bypass certain system conditions or to satisfy signer requirements~\cite{Maji11}. For example, a computer science student could collude with an engineering professor to sign a message as a computer science professor.

\subsection{Physical Signatures}

Physical signatures are signatures based on features of the physical world. These signatures are strong complements to cryptographic schemes in applications where there is constant interaction with the physical world~\cite{Wu17PCS}. These signatures can be either intrinsic or extrinsic, depending on how they are generated~\cite{Wu17PCS}. Intrinsic signatures are generated from inherent characteristics from devices, channels, or the physical environment~\cite{garg2013geo}. For instance, Wu {\em et al.}~\cite{garg2013geo} have proposed authenticating media through the Electric Network Frequency (ENF) of the place where the media was recorded. Extrinsic signatures, on the other hand, are signals and data that are intentionally injected and monitored in a system. For instance, physical-layer watermarks~\cite{satchidanandan2017dynamic}.


\section{Authentication for Pervasive Computing Applications} \label{sec:applied}

In this section, we highlight promising authentication methods for each of the application scenarios presented in Section~\ref{sec:scen}. Table~\ref{tab:match} summarizes what we consider to be interesting approaches for each scenario. Note, however, that these are based on the presented characteristics for each application scenario and authentication method. Naturally, even if an authentication method seems ill-suited for a particular application, it might be possible to come up with a viable authentication solution with it.

\begin{table}[t]
  \caption{Promising authentication methods for IoT application scenarios.}
  \label{tab:match}
  \centering
  \resizebox{\textwidth}{!}{%
  \begin{tabular}{c|c|c|c|c|c|c|c|c|c|c}
    \multicolumn{1}{c|}{} & {\bf ECDSA} & {\bf IBC} & {\bf Certificateless} & {\bf ABC}  & {\bf HBS}  & {\bf Group} & {\bf Ring} & {\bf Mesh} & {\bf Physical} & {\bf BLS}  \\ \hline
    WSN                &             &              &                       &            &            & \checkmark             &                       &                       &                           & \checkmark \\ \hline
    V2X                &             &              &                       & \checkmark &            &                        & \checkmark            & \checkmark            &                           &            \\ \hline
    Smart Homes        & \checkmark  &              &                       & \checkmark & \checkmark &                        &                       &                       &                           &            \\ \hline
    Wearables          &             & \checkmark   & \checkmark            &            & \checkmark &                        &                       &                       & \checkmark                & \checkmark \\ \hline
    CPHS               & \checkmark  &              &                       &            &            &                        &                       &                       & \checkmark                &            \\ 
  \end{tabular}%
  }
\end{table}

\subsection{Wireless Sensor Networks} \label{sec:applied.wsn}

The biggest impediment on authentication for WSNs are the inherent resource
constraint of the sensors. This means authentication solutions must rely on
methods that require low computing resources, low bandwidth, and low energy
consumption. These requirements, coupled with the presence of the resourceful
base station and previous bootstrapping process, make BLS an interesting alternative.
Likewise, group signatures are a natural fit for WSNs, with the base as the group manager.
However, the physical vulnerabilities of the sensors are a drawback for group signatures,
since compromising a single sensor would lead to the entire group being compromised. 
One way to mitigate this issue is to use secure hardware modules~\cite{Perrig02spins}.

\subsection{Vehicle-to-Everything}

There are several complications to V2X communications. For the most part, the major complication is the combination of high mobility, time-sensitive applications, and massive network size. This means vehicles must be able to very quickly authenticate several other vehicles in its vicinity. The added privacy requirements complicate design even further, as it means anonymous signature schemes are required.

Group signatures are not well suited for V2X applications, because the group manager would have to be one of the vehicles, which leads to trust concerns, and because the high mobility of vehicles would cause the group to change frequently. Ring signatures or mesh signatures are a better fit, since they do not require a group manager. Besides, vehicles would be able to sign messages as one of a group of vehicles in the vicinity, therefore protecting their privacy. Mesh signatures also have the added benefit of allowing threshold signatures, which would allow vehicles to group up to endorse information and naturally preventing adversaries from spreading false information. 

Another possible solution for V2X is ABC. In ABC, a vehicle could have a set of attributes that qualify it as a valid vehicle in the system, ensuring it can authenticate its own information without revealing its identity. ABC comes with the extra benefit of not allowing vehicles to collude and provides a simple framework to enable the deployment of vehicles with special characteristics, like emergency services, public transportation, or public service vehicles.

\subsection{Smart Homes}

The relaxed requirements of smart home environments facilitate the development of authentication solutions. Solutions need not be extremely efficient (except maybe for a handful of devices), networks are small, and applications are not time sensitive. Besides, the controlled environment and the direct involvement of users may enable novel solutions. HBS, is an interesting option for the resource-constrained devices of smart homes, especially since there are resourceful devices to handle the constant key generation. ABC is another interesting option, as it allows intuitive and fine-grained access control to devices, even among members of the household~\cite{Neto16}. IBC, Certificateless, and ECDSA are all also viable for smart homes, given the lax requirements of these applications, however, without the added benefits of HBS's high efficiency or ABC's fine-grained access control.


\subsection{Wearables}

The biggest challenge of authenticating wearable devices is their resource constraint and limited power supply. This means authentication solutions for these devices must be highly efficient. On the other hand, the periodic contact of these devices with other, more resourceful devices, enables solutions like HBS. The high efficiency of HBS makes them suitable for resource-constrained devices and the periodic contact with resourceful manager devices allow the keys to be easily renewed. Wearables also have frequent access to biometric or environmental information. This makes context-aware authentication solutions like physical signatures an interesting option. 

\subsection{CPHS}

It is complicated to propose a single authentication solution for CPHS given the very heterogeneous nature of its applications. But, there are a few staple features in CPHS system that hint at promising approaches. In particular, the constant involvement of the physical world makes physical signatures a viable, and interesting, authentication method. Naturally, this does not mean other authentication methods are ill-suited for CPHS. ABC may be interesting for automated workspaces, since access control can be implemented for each worker based on his attributes within the company hierarchy. More traditional approaches with simple certification schemes like IBC, certificateless or even ECDSA may also be viable where devices are not severely resource-constrained. For instance, in critical infrastructure systems where devices are resourceful and it is important to know who is signing messages for auditing purposes.
\section{Conclusion} \label{sec:conc}

Computing is quickly becoming more pervasive, creating a demand for stronger security solutions. In particular, authentication solutions tailored for pervasive applications are paramount, yet they remain an open challenge. In this paper, we reviewed noteworthy scenarios where pervasive computing is quickly establishing itself and indicated promising authentication approaches for each. We first identified the peculiarities of each application scenario as well as the main strengths and weakness of each authentication method. We then identified which authentication methods are well-suited for each scenario based on the peculiarities of each.


\bibliographystyle{splncs04}
\bibliography{ref}

\begin{thebibliography}{10}
\providecommand{\url}[1]{\texttt{#1}}
\providecommand{\urlprefix}{URL }
\providecommand{\doi}[1]{https://doi.org/#1}

\bibitem{Akkaya05}
Akkaya, K., Younis, M.: {A Survey on Routing Protocols for Wireless Sensor
  Networks}. Ad Hoc Networks  \textbf{3}(3),  325 -- 349 (2005)

\bibitem{Akyldiz10}
Akyildiz, I.F., Vuran, M.C.: Wireless sensor networks, vol.~4. John Wiley \&
  Sons (2010)

\bibitem{Riyami03}
Al-Riyami, S.S., Paterson, K.G.: {Certificateless Public Key Cryptography}. In:
  Asiacrypt (2003)

\bibitem{amadeo2016}
Amadeo, M., Campolo, C., Molinaro, A.: {Information-Centric Networking for
  Connected Vehicles: a Survey and Future Perspectives}. IEEE Communications
  Magazine  \textbf{54}(2),  98--104 (2016)

\bibitem{antonakakis2017understanding}
Antonakakis, M., April, T., Bailey, M., Bernhard, M., Bursztein, E., Cochran,
  J., Durumeric, Z., Halderman, J.A., Invernizzi, L., Kallitsis, M., et~al.:
  {Understanding the Mirai Botnet}. In: USENIX Security Symposium. pp.
  1092--1110 (2017)

\bibitem{Arampatzis05}
Arampatzis, T., Lygeros, J., Manesis, S.: {A Survey of Applications of Wireless
  Sensors and Wireless Sensor Networks}. In: Proceedings of the 2005 IEEE
  International Symposium on, Mediterranean Conference on Control and
  Automation Intelligent Control, 2005. (2005)

\bibitem{Atzori10}
Atzori, L., Iera, A., Morabito, G.: {The Internet of Things: A survey}.
  Computer networks  \textbf{54}(15),  2787--2805 (2010)

\bibitem{Bauza2010road}
Bauza, R., Gozalvez, J., Sanchez-Soriano, J.: Road traffic congestion detection
  through cooperative vehicle-to-vehicle communications. In: Local Computer
  Networks (LCN), 2010 IEEE 35th Conference on. pp. 606--612. IEEE (2010)

\bibitem{bernstein2009introduction}
Bernstein, D.J.: {Introduction to Post-quantum Cryptography}. Springer (2009)

\bibitem{boneh01}
Boneh, D., Franklin, M.K.: {Identity-Based Encryption from the Weil Pairing}.
  In: {International Cryptology Conference on Advances in Cryptology (CRYPTO)}
  (2001)

\bibitem{boneh2001short}
Boneh, D., Lynn, B., Shacham, H.: {Short Signatures from the Weil Pairing}.
  Advances in Cryptology—ASIACRYPT 2001 pp. 514--532 (2001)

\bibitem{boyen2007mesh}
Boyen, X.: {Mesh Signatures}. In: Conference on the Theory and Applications of
  Cryptographic Techniques (2007)

\bibitem{Brown01}
Brown, D.R., Gallant, R., Vanstone, S.A.: {Provably Secure Implicit Certificate
  Schemes}. In: International Conference on Financial Cryptography. pp.
  156--165. Springer (2001)

\bibitem{buchmann09}
Buchmann, J., Dahmen, E., Szydlo, M.: Hash-based digital signature schemes.
  Post-Quantum Cryptography pp. 35--93 (2009)

\bibitem{butin2015real}
Butin, D., Gazdag, S.L., Buchmann, J.: {Real-World Post-Quantum Digital
  Signatures}. In: Cyber Security and Privacy Forum (2015)

\bibitem{cardenas2008secure}
Cardenas, A.A., Amin, S., Sastry, S.: Secure control: Towards survivable
  cyber-physical systems. In: Distributed Computing Systems Workshops, 2008.
  ICDCS'08. 28th International Conference on. pp. 495--500. IEEE (2008)

\bibitem{chatterjee2011identity}
Chatterjee, S., Sarkar, P.: {Identity-Based Encryption}. Springer Science \&
  Business Media (2011)

\bibitem{chaum1991group}
Chaum, D., Van~Heyst, E.: {Group Signatures}. In: Advances in
  Cryptology—EUROCRYPT’91 (1991)

\bibitem{Derawi10}
Derawi, M.O., Nickel, C., Bours, P., Busch, C.: {Unobtrusive
  User-Authentication on Mobile Phones Using Biometric Gait Recognition}. In:
  Intelligent Information Hiding and Multimedia Signal Processing (IIH-MSP),
  2010 Sixth International Conference on. pp. 306--311. IEEE (2010)

\bibitem{DiPietro03}
Di~Pietro, R., Mancini, L.V.: {Security and Privacy Issues of Handheld and
  Wearable Wireless Devices}. Commun. ACM  \textbf{46}(9),  74--79 (2003)

\bibitem{Forster16}
Förster, D., Kargl, F., Löhr, H.: {PUCA: A Pseudonym Scheme with Strong
  Privacy Guarantees for Vehicular Ad-Hoc Networks}. Ad Hoc Networks
  \textbf{37},  122 -- 132 (2016)

\bibitem{galdi2017secure}
Galdi, C., Nappi, M., Dugelay, J.L.: {Secure User Authentication on Smartphones
  via Sensor and Face Recognition on Short Video Clips}. In: International
  Conference on Green, Pervasive, and Cloud Computing (GPC). pp. 15--22.
  Springer (2017)

\bibitem{garg2013geo}
Garg, R., Hajj-Ahmad, A., Wu, M.: {Geo-Location Estimation from Electrical
  Network Frequency Signals.} In: ICASSP (2013)

\bibitem{goyal2006attribute}
Goyal, V., Pandey, O., Sahai, A., Waters, B.: {Attribute-Based Encryption for
  Fine-Grained Access Control of Encrypted Data}. In: {Conference on Computer
  and Communications Security (CCS)} (2006)

\bibitem{hulsing2013w}
H{\"u}lsing, A.: {W-OTS+--Shorter Signatures for Hash-based Signature Schemes}.
  In: International Conference on Cryptology in Africa (2013)

\bibitem{jakubiak08}
Jakubiak, J., Koucheryavy, Y.: {State of the Art and Research Challenges for
  VANETs}. In: IEEE Consumer Communications and Networking Conference (2008)

\bibitem{Jing14}
Jing, Q., Vasilakos, A.V., Wan, J., Lu, J., Qiu, D.: {Security of the Internet
  of Things: Perspectives and Challenges}. Wireless Networks  \textbf{20}(8),
  2481--2501 (2014)

\bibitem{johnson2001elliptic}
Johnson, D., Menezes, A., Vanstone, S.: {The Elliptic Curve Digital Signature
  Algorithm (ECDSA)}. International Journal of Information Security
  \textbf{1}(1),  36--63 (2001)

\bibitem{karagiannis11}
Karagiannis, G., Altintas, O., Ekici, E., Heijenk, G., Jarupan, B., Lin, K.,
  Weil, T.: {Vehicular Networking: A Survey and Tutorial on Requirements,
  Architectures, Challenges, Standards and Solutions}. IEEE communications
  surveys \& tutorials  \textbf{13}(4),  584--616 (2011)

\bibitem{kargl2015wireless}
Kargl, F., Waldschmidt, C., Moser, S., Slomka, F., et~al.: {Wireless
  Channel-Based Message Authentication}. In: Vehicular Networking Conference
  (VNC) (2015)

\bibitem{Karlof04}
Karlof, C., Sastry, N., Wagner, D.: {TinySec: A Link Layer Security
  Architecture for Wireless Sensor Networks}. In: International Conference on
  Embedded Networked Sensor Systems (SenSys) (2004)

\bibitem{Kaur17game}
Kaur, R., Kaur, N., Sood, S.K.: {Security in IoT Network Based on Stochastic
  Game Net Model}. International Journal of Network Management  (2017)

\bibitem{lee2008cyber}
Lee, E.A.: {Cyber Physical Systems: Design Challenges}. In: Object oriented
  real-time distributed computing (isorc), 2008 11th ieee international
  symposium on. pp. 363--369 (2008)

\bibitem{perrig06seven}
Luk, M., Perrig, A., Whillock, B.: {Seven Cardinal Properties of Sensor Network
  Broadcast Authentication}. In: {Workshop on Security of Ad Hoc and Sensor
  Networks (SASN)} (2006)

\bibitem{Maji11}
Maji, H.K., Prabhakaran, M., Rosulek, M.: {Attribute-Based Signatures}. In:
  CT-RSA (2011)

\bibitem{menezes1996handbook}
Menezes, A.J., Van~Oorschot, P.C., Vanstone, S.A.: {Handbook of Applied
  Cryptography}. CRC press (1996)

\bibitem{Neto16}
Neto, A.L.M., Souza, A.L., Cunha, I., Nogueira, M., Nunes, I.O., Cotta, L.,
  Gentille, N., Loureiro, A.A., Aranha, D.F., Patil, H.K., Oliveira, L.B.:
  {AoT: Authentication and Access Control for the Entire IoT Device
  Life-Cycle}. In: SenSys (2016)

\bibitem{oliveira2011secure}
Oliveira, L.B., Kansal, A., Gouv{\^e}a, C.P., Aranha, D.F., L{\'o}pez, J.,
  Priyantha, B., Goraczko, M., Zhao, F.: {Secure-TWS: Authenticating Node to
  Multi-User Communication in Shared Sensor Networks}. The Computer Journal
  \textbf{55}(4),  384--396 (2011)

\bibitem{ouechtati2017trust}
Ouechtati, H., Azzouna, N.B.: {Trust-ABAC Towards an Access Control System for
  the Internet of Things}. In: International Conference on Green, Pervasive,
  and Cloud Computing (GPC). pp. 75--89. Springer (2017)

\bibitem{Perrig02spins}
Perrig, A., Szewczyk, R., Tygar, J.D., Wen, V., Culler, D.E.: {SPINS: Security
  Protocols for Sensor Networks}. Wireless networks  \textbf{8}(5),  521--534
  (2002)

\bibitem{qian2008design}
Qian, Y., Moayeri, N.: Design of secure and application-oriented vanets. In:
  Vehicular Technology Conference, 2008. VTC Spring 2008. IEEE. pp. 2794--2799.
  IEEE (2008)

\bibitem{Raij11}
Raij, A., Ghosh, A., Kumar, S., Srivastava, M.: {Privacy Risks Emerging from
  the Adoption of Innocuous Wearable Sensors in the Mobile Environment}. In:
  Proceedings of the SIGCHI Conference on Human Factors in Computing Systems
  (2011)

\bibitem{Raya07}
Raya, M., Hubaux, J.P.: Securing vehicular ad hoc networks. Journal of Computer
  Security  \textbf{15}(1),  39--68 (2007)

\bibitem{rivest2001leak}
Rivest, R., Shamir, A., Tauman, Y.: {How to Leak a Secret}. Advances in
  Cryptology—ASIACRYPT 2001  (2001)

\bibitem{rohde08}
Rohde, S., Eisenbarth, T., Dahmen, E., Buchmann, J., Paar, C.: Efficient
  hash-based signatures on embedded devices. SECSI-Secure Component and System
  Identification, Berlin, Germany  (2008)

\bibitem{Rostami13}
Rostami, M., Juels, A., Koushanfar, F.: {Heart-to-Heart (H2H): Authentication
  for Implanted Medical Devices}. In: {CCS} (2013)

\bibitem{satchidanandan2017dynamic}
Satchidanandan, B., Kumar, P.: Dynamic watermarking: Active defense of
  networked cyber--physical systems. Proceedings of the IEEE  \textbf{105}(2),
  219--240 (2017)

\bibitem{satyanarayanan2001pervasive}
Satyanarayanan, M., et~al.: Pervasive computing: Vision and challenges. IEEE
  Personal Communications  \textbf{8}(4),  10--17 (2001)

\bibitem{schirner2013future}
Schirner, G., Erdogmus, D., Chowdhury, K., Padir, T.: {The Future of
  Human-in-the-Loop Cyber-Physical Systems}. Computer  \textbf{46}(1),  36--45
  (2013)

\bibitem{shamir84}
Shamir, A.: {Identity-based Cryptosystems and Signature Schemes}. In:
  {International Cryptology Conference on Advances in Cryptology (CRYPTO)}
  (1984)

\bibitem{Souza17Nomadikey}
Souza, A., Cunha, {\'I}., B~Oliveira, L.: {NomadiKey: User Authentication for
  Smart Devices based on Nomadic Keys}. International Journal of Network
  Management pp. e1998--n/a (2017)

\bibitem{Stinson02}
Stinson, D.: {Cryptography: Theory and Practice}. CRC/C\&H (2002)

\bibitem{Wei14}
Wei, J.: {How Wearables Intersect with the Cloud and the Internet of Things:
  Considerations for the Developers of Wearables.} Consumer Electronics
  Magazine  \textbf{3}(3),  53--56 (2014)

\bibitem{weiss2011}
Wei{\ss}, C.: {V2X Communication in Europe--From Research Projects Towards
  Standardization and Field Testing of Vehicle Communication Technology}.
  Computer Networks  \textbf{55}(14),  3103--3119 (2011)

\bibitem{Wu17PCS}
Wu, M., Quintão~Pereira, F., Liu, J., Ramos, H., Alvim, M., Oliveira, L.: {New
  Directions: Proof-Carrying Sensing –- Towards Real-World Authentication in
  Cyber-Physical Systems}. In: Conference on Embedded Networked Sensor Systems
  (SenSys) (2017)

\bibitem{Ye02}
Ye, W., Heidemann, J., Estrin, D.: {An Energy-Efficient MAC Protocol for
  Wireless Sensor Networks}. In: Proceedings.Twenty-First Annual Joint
  Conference of the IEEE Computer and Communications Societies (2002)

\bibitem{yuan2005}
Yuan, E., Tong, J.: {Attributed Based Access Control (ABAC) for Web Services}.
  In: {International Conference on Web Services (ICWS)} (2005)

\end{thebibliography}

\end{document}